%% file: main.tex
\documentclass[a4paper,11pt]{article}
\usepackage{pos}
\usepackage{tikz}
\usepackage{mathtools}
\usepackage{subcaption}
\usepackage{float}
\floatstyle{plaintop}
\restylefloat{table}
\usepackage{caption}
\usepackage[force]{feynmp-auto}
\usepackage{tabularx}
\usepackage{multirow, makecell}

\renewcommand{\>}{\rangle}
\newcommand{\<}{\langle}
\newcommand{\bomega}{\bar{\omega}}

\title{Generalised Parton Distributions from Lattice Feynman-Hellmann Techniques}

\manuallySeparateAuthors
\author*[a]{A.~Hannaford-Gunn,}
\author[a]{ K.~U.~Can,}
    \author[b]{ R.~Horsley,}
    \author[c]{ H.~Perlt,}
\author[d]{ P.~E.~L.~Rakow,}
    \author[e]{ G.~Schierholz,}
      \author[f]{ H.~St\"uben,}
    \author[a]{ R.~D.~Young}
    \author[a]{ and J.~M.~Zanotti}
    \author{ for the {CSSM-QCDSF-UKQCD Collaboration}}

\affiliation[a]{CSSM, Department of Physics, University of Adelaide,
  Adelaide SA 5005, Australia}
  
\affiliation[b]{School of Physics, University of Edinburgh, Edinburgh EH9 3JZ, UK}
\affiliation[c]{Insitut f\"ur Theoretische Physik, Universist\"at Leipzig, 04103 Leipzig, Germany}

\affiliation[d]{Theoretical Physics Division, Department of Mathematical Sciences, University of Liverpool, Liverpool L69 3BX, UK}

    \affiliation[e]{Deutsches Elektronen-Synchrotron DESY, Notkestr.~85, 22607 Hamburg, Germany}
     
        \affiliation[f]{Regionales Rechenzentrum, Universit\"at Hamburg, 20146 Hamburg, Germany}

\emailAdd{alec.hannafordgunn@adelaide.edu.au}

\abstract{
We report on the use of Feynman-Hellmann techniques to calculate the off-forward Compton amplitude (OFCA) in lattice QCD. At leading-twist, the Euclidean OFCA is parameterised by the Mellin moments of generalised parton distributions (GPDs). Hence we extract GPD moments for two values of the soft momentum transfer, $t=-1.10, -2.20\;\text{GeV}^2$ and zero-skewness kinematics at an unphysical pion mass of $m_{\pi}\approx 470\;\text{MeV}$. This includes the first determination of the $n=4$ moments.
}

\FullConference{%
 The 38th International Symposium on Lattice Field Theory, LATTICE2021
  26th-30th July, 2021
  Zoom/Gather@Massachusetts Institute of Technology
}
\numberwithin{equation}{section}

\begin{document}
\maketitle

\input{section1}

\input{section2}

\input{section3}
\input{section4}

\input{conclusion}
\input{acknowledgements}

\bibliographystyle{JHEP}
\bibliography{refer}

\end{document}

%% file: section1.tex
\section{Introduction}

Generalised parton distributions (GPDs) \cite{mullerscaling, jiog, radyushkinscaling} are observables that contain a staggering amount of hadronic information, including the spatial distribution \cite{burkardt} and spin structure \cite{jiog} of constituent quarks and gluons, and the pressure distributions within hadrons \cite{dtermexp}. However, experimental probes of GPDs are fraught with difficulties. In particular, global fits require assumptions about the functional form of GPDs that are beyond our current understanding \cite{gpdphenomreview}. For this reason, there has been strong interest in lattice QCD studies of GPDs. Historically, lattice studies have been limited to their lowest Mellin moments; the highest calculated so far are the $n=3$ moments \cite{gpdlatt1, gpdlatt2, gpdlatt4, gpdlatt5, gpdlatt6}. More recently, there has been a great deal of interest in calculating parton distributions from equal-time, non-local correlators in lattice QCD \cite{radpseudo, jiquasi}, including calculations of quasi-GPDs \cite{pionquasi, nucleonquasi1, nucleonquasi2}.

\begin{figure}[t!]
\centering
\includegraphics[width=0.48\textwidth]{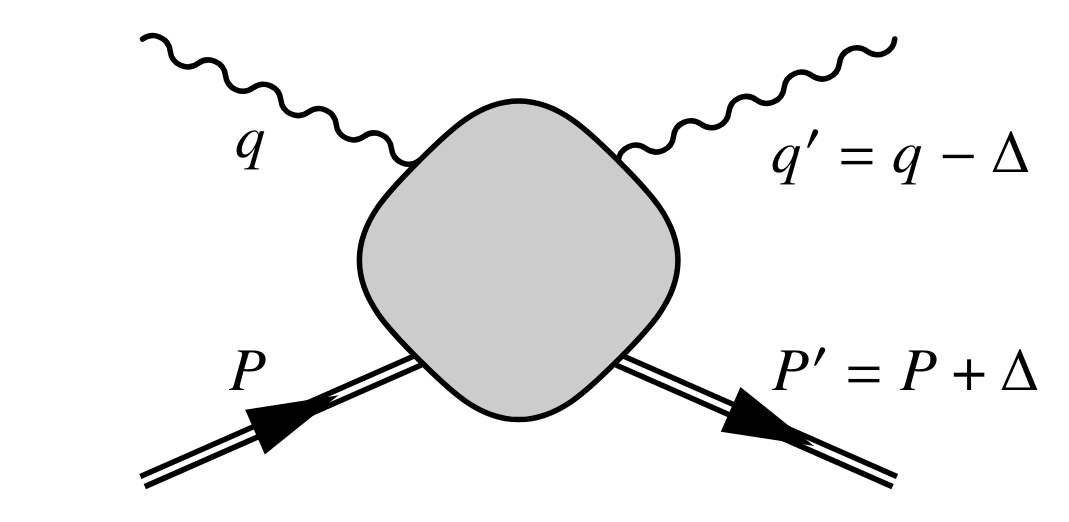}
    \vspace{3mm}
    \caption{The Feynman diagram for off-forward Compton scattering $\gamma^{*} (q)N(P)  \to \gamma^{*}(q')N(P') $.}
    \label{offfwdpic}
    \end{figure}

Here, we report on a lattice QCD calculation of the off-forward Compton amplitude (OFCA),
\begin{equation}
    T^{\mu\nu}\equiv i\int d^4ze^{\frac{i}{2}(q+q')\cdot z} \<P'|T\{j^{\mu}(z/2)j^{\nu}(-z/2)\}|P\>,
    \label{vvcadef}
\end{equation}
which describes the process of nucleon-photon scattering: $\gamma^{*} (q)N(P)  \to \gamma^{*}(q')N(P') $, with $q_{\mu}\neq q'_{\mu}$ (see Figure \ref{offfwdpic}). At high energies ($|q^2|$ and/or $|q'^2|\gg \Lambda_{\text{QCD}}^2$), this amplitude is dominated by a convolution of GPDs \cite{jiog}. Therefore, we can use a lattice calculation of this amplitude to determine GPD-related quantities.

The method we use to calculate the OFCA is an extension of Feynman-Hellmann methods, which have previously been used to calculate the \emph{forward} Compton amplitude \cite{fwdletter, comptonproceedings, fwdpaper, utkucallangross, interlacingsubtraction}, and off-forward elastic form factors \cite{collabffs}. This involves computing nucleon propagators in the presence of weakly-coupled background fields. By isolating the contribution that is quadratic in this coupling, we can calculate four-point functions. As such, Feynman-Hellmann methods provide a realistic alternative to the direct computation of four-point functions.

The numerical results presented here are at the SU(3) flavour symmetric point and a larger-than-physical pion mass \cite{configs}. In terms of kinematics, we are at the zero-skewness point, which is not accessible to experiment but is the limit in which GPDs encode spatial distributions of quarks \cite{burkardt}. Moreover, we calculate two values of the soft momentum transfer, $t=-1.10, -2.20 \;\text{GeV}^2$, with a hard momentum transfer of $\bar{Q}^2 \approx 6-7\;\text{GeV}^2$. 

For this preliminary calculation, we consider this hard scale sufficiently large to assume that the extracted amplitude is dominated by its GPD contributions. Therefore, we also present Mellin moment fits, which we interpret as GPD moments. This includes the first determination of the $n=4$ moments. A more detailed discussion of the work presented here can be found in Ref.~\cite{offfwdpaper}.

%% file: section2.tex
\section{Feynman-Hellmann Methods}

In this section we will give a brief derivation of the Feynman-Hellmann relation that allows us to access the OFCA. We start with the perturbed quark propagators that we calculate:
\begin{equation}
        \begin{split}
            S_{\vec{\lambda}}  = \big [ \underbrace{M}_{\text{fermion matrix}} - \underbrace{\lambda_1\mathcal{J}_3 (\vec{q}_1)- \lambda_2\mathcal{J}_3 (\vec{q}_2)}_{\text{background fields}}\big]^{-1} &  = \underbrace{M^{-1}}_{\text{unperturbed}} + \sum_i\lambda_i \underbrace{M^{-1}\mathcal{J}_3(\vec{q}_i) M^{-1}}_{\text{three-point}} 
       \\ & + \sum_{i,j}\lambda_i\lambda_j \underbrace{M^{-1}\mathcal{J}_3 (\vec{q}_i)M^{-1}\mathcal{J}_3(\vec{q}_j) M^{-1}}_{\text{four-point}} + \dots
        \label{pertprop}
        \end{split}
    \end{equation}
    Here, our couplings, $\lambda_{1,2}$, are small, and $\vec{q}_1\neq\vec{q}_2$ are our inserted momenta. We choose our perturbing matrices to be $[\mathcal{J}_3(\vec{q}_j)]_{n,m} = \delta_{x_n,x_m}2\cos(\vec{q}_j\cdot\vec{x}_n)i\gamma_3$.
    
    Taking a mixed, second-order derivative gives 
    \begin{equation}
        \frac{\partial^2}{\partial\lambda_1\partial\lambda_2}S_{\vec{\lambda}}\bigg |_{\vec{\lambda}=0} = {M^{-1}\mathcal{J}_3 ({\vec{q}_1})M^{-1}\mathcal{J}_3({\vec{q}_2}) M^{-1}} + (1\leftrightarrow 2),
        \label{quarkpropderiv}
    \end{equation}
    which is a four-point function with momentum transfer.
    
    We can insert these quark propagators either as up or down quarks into a nucleon propagator:
    \begin{equation*}
    \mathcal{G}^d_{\vec{\lambda}} \simeq \big\< S^u S^u S^d_{\vec{\lambda}}\big\>, \quad \mathcal{G}^u_{\vec{\lambda}} \simeq \big\< S^u_{\vec{\lambda}} S^u_{\vec{\lambda}} S^d\big\>,
    \end{equation*}
    where we have suppressed the spin and flavour structure of the nucleon propagators.
    

\begin{figure}
    \centering
    \includegraphics[width=\textwidth]{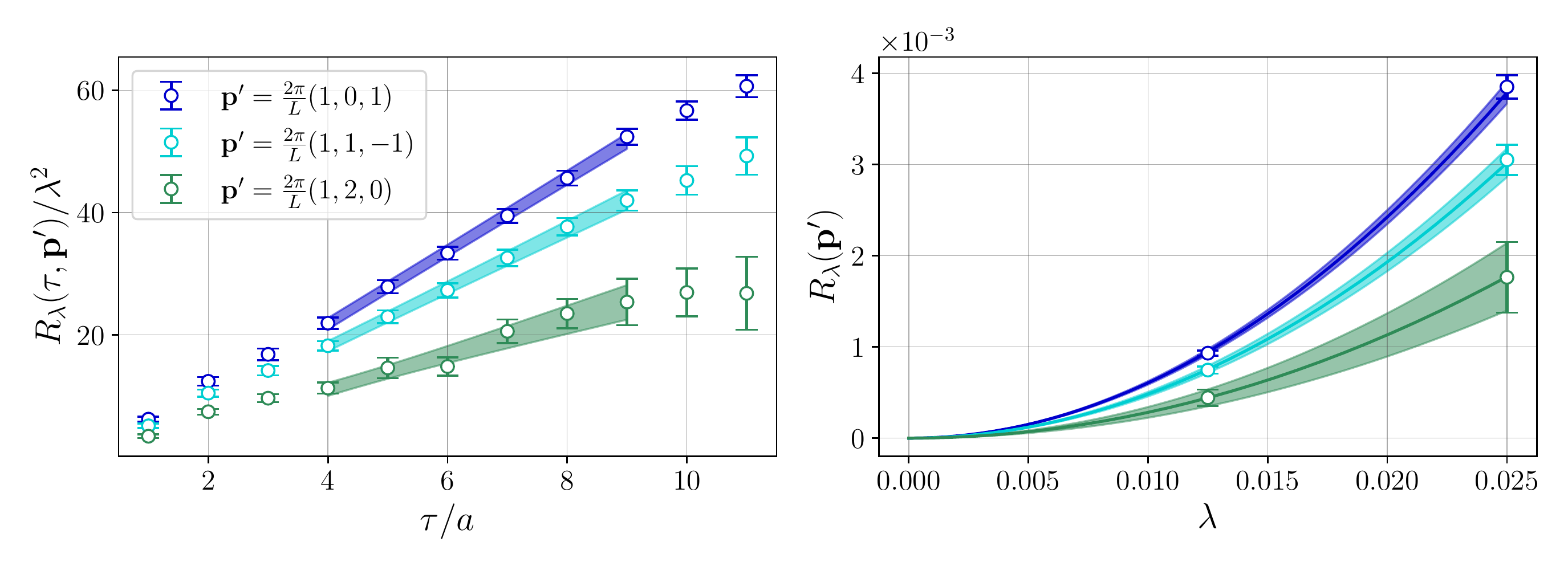}
    \caption{Left: the Euclidean time dependence of the ratio defined in Eq.~\eqref{combocorr}, with a linear fit $f(\tau) = a\tau +b$. Right: after fitting in Euclidean time, we demonstrate that the $\lambda^2$ term is dominant by fitting $g(\lambda) = b\lambda^2$.}
    \label{fig:taulambdadep}
\end{figure}
    
    Then, as in Eq.~\eqref{quarkpropderiv}, we can take a mixed, second-order derivative to get
    \begin{equation}
         \begin{split}
        & \frac{\partial^2}{\partial\lambda_1\partial\lambda_2}\frac{\mathcal{G}_{\vec{\lambda}} (\tau, \vec{p}')}{\mathcal{G}_{0} (\tau, \vec{p}')}\bigg|_{\vec{\lambda}=0}
       \simeq
       \frac{\tau}{2E_N(\vec{p}')} \frac{\sum_{s',s}\text{tr}\big[\Gamma u(\vec{p}',s')T_{33}(\vec{p}';\vec{q}_1, \vec{q}_2)\bar{u}(\vec{p},s)\big]}{\sum_{s}\text{tr}\big[\Gamma u(\vec{p}',s)\bar{u}(\vec{p}',s)\big]},
       \label{fhref}
    \end{split}
    \end{equation}
    where 
    \begin{equation*}
        T_{\mu\nu}(\vec{p}';\vec{q}_1, \vec{q}_2) = \sum_{z}e^{\frac{i}{2}(\vec{q}_1+\vec{q}_2)\cdot \vec{z}}\<N(\vec{p}')|T\{j_{\mu}(z)j_{\nu}(0)\}|N(\vec{p}'-\vec{q}_1+\vec{q}_2)\>,
    \end{equation*}
    a discretisation of the OFCA, Eq.~\eqref{vvcadef}. Note that a more complete derivation of Eq.~\eqref{fhref} is presented in Ref.~\cite{offfwdpaper}.
    
    We approximate the mixed, second-order derivative with the ratio
    \begin{equation}
    R_{\lambda} \equiv \frac{\mathcal{G}_{(\lambda,\lambda)}+\mathcal{G}_{(-\lambda,-\lambda)}-\mathcal{G}_{(\lambda,-\lambda)}-\mathcal{G}_{(-\lambda,\lambda)}}{\mathcal{G}_{(0,0)}},
    \label{combocorr}
\end{equation}
and use a linear fit in Euclidean time, $\tau$, to extract the OFCA (Fig.~\ref{fig:taulambdadep}). 

After fitting in Euclidean time, we can fit $R_{\lambda}$ across multiple $\lambda$ to a quadratic function, $g(\lambda) = b\lambda^2$, as shown in Fig.~\ref{fig:taulambdadep}. The results are well-described by a quadratic, which confirms that we are extracting the $\lambda_1\lambda_2$ contribution that is proportional to the OFCA.

%% file: section3.tex
\section{Parameterisation of the Compton amplitude}

In the previous section, we outlined a method to calculate the OFCA in lattice QCD. Now, we briefly discuss how to parameterise the OFCA in terms of GPD moments.

To begin, we define four linearly independent Lorentz scalars that our OFCA is a function of:
 \begin{equation}
    \bomega=-\frac{2(P+P')\cdot(q+q')}{(q+q')^2}, \quad  \xi=\frac{q'^2 -q^2}{(P+P')\cdot(q+q')}, \quad
    t = (P'-P)^2, \quad \bar{Q}^2 = -\frac{1}{4}(q+q')^2.
    \label{scalardef}
\end{equation}

It is well-known that, for large $\bar{Q}^2$, the off-forward Compton amplitude is dominated by contributions from GPDs \cite{jiog}:
\begin{equation*}
\begin{split}
    T^{\mu\nu} (\bomega,\xi,t, \bar{Q}^2)\simeq g^{\mu\nu}\bomega^2\int dx  \frac{x  G(x,\xi,t)}{1 -x^2\bomega^2 -i\epsilon} +\dots +\mathcal{O}\big({1}/{\bar{Q}^2}\big),
\end{split}
\end{equation*} 
where $G$ is a GPD. Or in the Euclidean region, $|\bomega|<1$,
\begin{equation*}
\begin{split}
    T^{\mu\nu} (\bomega,\xi,t, \bar{Q}^2)\simeq g^{\mu\nu}\sum_n \bomega ^n\int dx x^{n-1} G(x,\xi,t) +\dots+\mathcal{O}\big({1}/{\bar{Q}^2}\big).
\end{split}
\end{equation*} 


A complete leading-twist operator product expansion (OPE) of the OFCA with leading-order Wilson coefficients has been calculated, and is presented in Ref.~\cite{offfwdpaper}. Here, we will present the final result of that work, which is relevant to interpreting the lattice results.

From Eq.~\eqref{fhref}, we can see that the quantity of interest is 
\begin{equation}
     \mathcal{R}(\bomega, t,\bar{Q}^2)\equiv \frac{\sum _{s,s'} \text{tr}\big[\Gamma u(P',s')T_{33}  \bar{u}(P,s)\big]}{\sum _{s}{\text{tr}[\Gamma u(P',s)\bar{u}(P',s)]}}.
    \label{lattcompton}
\end{equation}

First, we note a few extra conditions we apply to our numerical results
\begin{enumerate}
    \item We use zero-skewness kinematics ($\xi=0$). From Eq.~\eqref{scalardef}, this is equivalent to $\vec{q}_1^2 = \vec{q}_2^2$.
    \item We use the spin-parity projector $\Gamma = \frac{1}{2}(\mathbb{I} + \gamma_4)$.
\item We subtract off the $\bomega=0$ contribution: $\overline{\mathcal{R}}(\bomega, t, \bar{Q}^2) = \mathcal{R}(\bomega, t,\bar{Q}^2)- \mathcal{R}(\bomega=0, t,\bar{Q}^2)$.
\end{enumerate}

The final parameterisation we fit to is then
\begin{equation}
    \begin{split}
        \overline{\mathcal{R}}^q(\bomega,  t,\bar{Q}^2) = 
        2K_{33} \sum_{n=2,4,6}^{\infty}\bomega ^{n}M^q_{n}(t),
         \label{nucleonLOFH}
    \end{split}
\end{equation}
where $K_{33}$ is a kinematic factor we can divide out, and we define 
\begin{equation}
    \begin{split}
        M^q_{n}(x,t) \equiv \int _{-1} ^1 dxx^{n-1} \bigg[H^q(x,t) +\frac{t}{8m_N^2} E^q(x,t) \bigg],
        \label{momcombo}
    \end{split}
\end{equation}
the moments of a linear combination of the unpolarised GPDs at zero-skewness. Calculating the independent contributions of $H$ and $E$ GPD moments is a goal of future work.

%% file: section4.tex
\section{Results: Moment Fits and Compton Amplitude}

		\begin{table*}
				\centering
				\caption{ \label{tab:gauge_details} Details of the gauge ensemble used in this work.}
				\setlength{\extrarowheight}{2pt}
	    	\begin{tabularx}{\textwidth}{ccccccccc}
					\hline\hline
					$N_f$  & $\kappa_l$ & $\kappa_s$ & $L^3 \times T$ & $a$ & $m_\pi$  & $m_\pi L$ & $Z_V$ & $N_\text{cfg}$\\
					&&&& [fm] &[GeV]&&& \\
					\hline
					$2+1$ & 0.1209  & 0.1209  & $32^3\times64$ & 0.074(2) & $0.467(12)$ & $\sim 5.6$ & 0.8611(84) & 1763 \\
					\hline\hline
				\end{tabularx}
			\end{table*}
			
			Details of the gauge ensembles used in this work are given in Tab.~\ref{tab:gauge_details}. We calculate two sets of perturbed correlators, with two pairs of inserted momenta, $\vec{q}_{1,2}$:
        \begin{table}[h!]
            \centering
			\begin{tabularx}{.63\textwidth}{c|c|c|c|c}
				\hline\hline
			 	Set & 	$\frac{L}{2\pi}\vec{q}_1$,	$\frac{L}{2\pi}\vec{q}_2$ & $t\;[\text{GeV}^2]$ &  $\bar{Q}^2\;[\text{GeV}^2]$ &$N_{\text{meas}}$ 	\\
				\hline
				\multirowcell{1}{\#1} 
			                        &	$(1,5,1)$, $(-1,5,1)$ & $-1.10$ & $7.13$ & $996$ \\ 
				\hline
				\multirowcell{1}{\#2} 
			                        &	 $(4,2,2)$, $(2,4,2)$ & $-2.20$ & $6.03$ & $996$ \\ 
				\hline
				\hline
			\end{tabularx}
        \end{table}
        
			We then vary the $\bomega$ variable by varying our sink momentum $\vec{p}'$. Since $|\bomega|<1$, we can truncate the expansion in $\bomega$, Eq.~\eqref{nucleonLOFH}, at some power $J$ for $\bomega^{2J}$. Then, using Markov chain Monte Carlo methods \cite{pymc3_no1, pymc3_no2}, we can fit the moments defined in Eq.~\eqref{momcombo}. We assume monotonically decreasing moments:
			\begin{equation*}
			    M_{2}(t)\geq M_4(t) \geq ... \geq M_{2J}(t).
			\end{equation*}
			However, future work will aim to incorporate model-independent constraints on GPDs \cite{Pobylitsa_2002, Pobylitsa_2002_IPS, Pobylitsa_2004} and on the Compton amplitude \cite{compton_positivity} to derive better prior conditions.
			
			In our case, we fit the first four moments, $n=2,4,6,8$, and report the first two. 
\begin{figure}
    \centering
    \includegraphics[width=\textwidth]{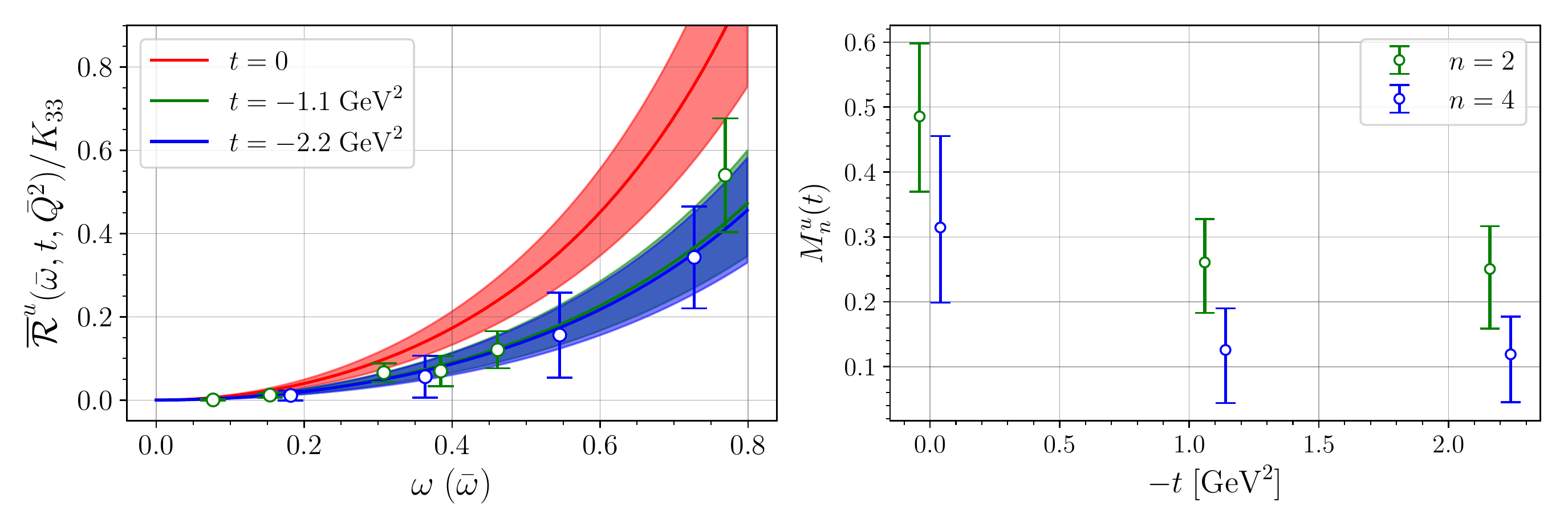}
    \caption{Left: the quantity defined in Eq.~\eqref{nucleonLOFH}, $\overline{\mathcal{R}}$, for up quarks, divided by a kinematic factor. Right: the moments, $M_{n}$, defined in Eq.~\eqref{momcombo} fit from $\overline{\mathcal{R}}$ for up quarks.}
    \label{fig:moments_omega}
\end{figure}
			Plots of $\overline{\mathcal{R}}$ and the extracted moments are given in Fig.~\ref{fig:moments_omega}, where we observe that the values of $\overline{\mathcal{R}}$ from our lattice simulation are well described by the parameterisation with the moments. Moreover, the $M_2$ moments are consistent with those calculated using three-point methods \cite{gpdlatt6}. The $M_4$ moments have never been calculated before, and hence this study is a first look at the $t$ behaviour of these moments. We observe a decrease with $-t$ for both moments, as expected. At present, however, the results are too noisy and exploratory to draw strong distinctions between the $t$ dependence of the two moments.
	

%% file: conclusion.tex
\section{Conclusion and Outlook}

In these proceedings, we have reported on the determination of the off-forward Compton amplitude in lattice QCD. This calculation employed an extension of Feynman-Hellmann methods that have previously been applied to numerous matrix elements, including the forward Compton amplitude.

Although this calculation is highly exploratory, the initial results are very promising. Future work will be aimed at:
\begin{enumerate}
    \item Controlling the systematic errors, including higher-twist corrections and the anomalous asymptotic behaviour of the subtraction function \cite{interlacingsubtraction}.
    \item Separating out the moments of the helicity-conserving and -flipping GPDs, $H$ and $E$, respectively (see Eq.~\eqref{momcombo}).
    \item Calculating a greater kinematic spread of $t$ and $\bar{Q}^2$ values.
\end{enumerate}
This will allow us to fit many more GPD moments, and report their higher-twist contributions more accurately. Moreover, it would allow us to constrain GPD models, and apply other methods to access GPDs directly from the Euclidean OFCA \cite{montecarlofit}.

%% file: acknowledgements.tex
\section{Acknowledgements}

The numerical configuration generation (using the BQCD lattice QCD program~\cite{Haar:2017ubh})) and data analysis (using the Chroma software library~\cite{Edwards:2004sx}) was carried out on the DiRAC Blue Gene Q and Extreme Scaling (EPCC, Edinburgh, UK) and Data Intensive (Cambridge, UK) services, the GCS supercomputers JUQUEEN and JUWELS (NIC, Jülich, Germany) and resources provided by HLRN (The North-German Supercomputer Alliance), the NCI National Facility in Canberra, Australia (supported by the Australian Commonwealth Government) and the Phoenix HPC service (University of Adelaide). AHG is supported by an Australian Government Research Training Program (RTP) Scholarship. RH is supported by STFC through grant ST/P000630/1. PELR is supported in part by the STFC under contract ST/G00062X/1. GS is supported by DFG Grant No. SCHI 179/8-1. KUC, RDY and JMZ are supported by the Australian Research Council grant DP190100297. 

%% file: main.bbl
\providecommand{\href}[2]{#2}\begingroup\raggedright\begin{thebibliography}{10}

\bibitem{mullerscaling}
D.~M\"uller, D.~Robaschik, B.~Geyer, F.M.~Dittes and J.~Ho\v{r}ej\v{s}i,
  \emph{{Wave functions, evolution equations and evolution kernels from light
  ray operators of QCD}},
  \href{https://doi.org/10.1002/prop.2190420202}{\emph{Fortsch. Phys.}
  {\bfseries 42} (1994) 101}
  [\href{https://arxiv.org/abs/hep-ph/9812448}{{\ttfamily hep-ph/9812448}}].

\bibitem{jiog}
X.-D.~Ji, \emph{{Gauge-Invariant Decomposition of Nucleon Spin}},
  \href{https://doi.org/10.1103/PhysRevLett.78.610}{\emph{Phys. Rev. Lett.}
  {\bfseries 78} (1997) 610}
  [\href{https://arxiv.org/abs/hep-ph/9603249}{{\ttfamily hep-ph/9603249}}].

\bibitem{radyushkinscaling}
A.V.~Radyushkin, \emph{{Nonforward parton distributions}},
  \href{https://doi.org/10.1103/PhysRevD.56.5524}{\emph{Phys. Rev. D}
  {\bfseries 56} (1997) 5524}
  [\href{https://arxiv.org/abs/hep-ph/9704207}{{\ttfamily hep-ph/9704207}}].

\bibitem{burkardt}
M.~Burkardt, \emph{{Impact parameter dependent parton distributions and off
  forward parton distributions for zeta ---\ensuremath{>} 0}},
  \href{https://doi.org/10.1103/PhysRevD.62.071503}{\emph{Phys. Rev. D}
  {\bfseries 62} (2000) 071503}
  [\href{https://arxiv.org/abs/hep-ph/0005108}{{\ttfamily hep-ph/0005108}}].

\bibitem{dtermexp}
V.D.~Burkert, L.~Elouadrhiri and F.X.~Girod, \emph{The pressure distribution
  inside the proton}, {\emph{Nature} {\bfseries 557} (2018) 396}.

\bibitem{gpdphenomreview}
K.~Kumericki, S.~Liuti and H.~Moutarde, \emph{{GPD phenomenology and DVCS
  fitting}: {Entering the high-precision era}},
  \href{https://doi.org/10.1140/epja/i2016-16157-3}{\emph{Eur. Phys. J. A}
  {\bfseries 52} (2016) 157}
  [\href{https://arxiv.org/abs/1602.02763}{{\ttfamily 1602.02763}}].

\bibitem{gpdlatt1}
P.~H{\"a}gler, J.W.~Negele, D.B.~Renner, W.~Schroers, T.~Lippert and
  K.~Schilling, \emph{{Moments of nucleon generalized parton distributions in
  lattice QCD}}, \href{https://doi.org/10.1103/PhysRevD.68.034505}{\emph{Phys.
  Rev. D} {\bfseries 68} (2003) 034505}
  [\href{https://arxiv.org/abs/hep-lat/0304018}{{\ttfamily hep-lat/0304018}}].

\bibitem{gpdlatt2}
M.~G{\"o}ckeler, R.~Horsley, D.~Pleiter, P.E.L.~Rakow, A.~Sch{\"a}fer,
  G.~Schierholz et~al., \emph{{Generalized parton distributions from lattice
  QCD}}, \href{https://doi.org/10.1103/PhysRevLett.92.042002}{\emph{Phys. Rev.
  Lett.} {\bfseries 92} (2004) 042002}
  [\href{https://arxiv.org/abs/hep-ph/0304249}{{\ttfamily hep-ph/0304249}}].

\bibitem{gpdlatt4}
M.~G\"ockeler, P.~H{\"a}gler, R.~Horsley, Y.~Nakamura, D.~Pleiter, P.E.L.~Rakow
  et~al., \emph{{Transverse spin structure of the nucleon from lattice {QCD}
  simulations}},
  \href{https://doi.org/10.1103/PhysRevLett.98.222001}{\emph{Phys. Rev. Lett.}
  {\bfseries 98} (2007) 222001}
  [\href{https://arxiv.org/abs/hep-lat/0612032}{{\ttfamily hep-lat/0612032}}].

\bibitem{gpdlatt5}
M.~Ohtani, D.~Br\"ommel, M.~G\"ockeler, P.~H{\"a}gler, R.~Horsley, Y.~Nakamura
  et~al., \emph{{Moments of generalized parton distributions and quark angular
  momentum of the nucleon}},
  \href{https://doi.org/10.22323/1.042.0158}{\emph{PoS} {\bfseries LATTICE2007}
  (2007) 158} [\href{https://arxiv.org/abs/0710.1534}{{\ttfamily 0710.1534}}].

\bibitem{gpdlatt6}
P.~Hägler, W.~Schroers, J.~Bratt, J.W.~Negele, A.V.~Pochinsky, R.G.~Edwards
  et~al., \emph{{Nucleon Generalized Parton Distributions from Full Lattice
  QCD}}, \href{https://doi.org/10.1103/PhysRevD.77.094502}{\emph{Phys. Rev. D}
  {\bfseries 77} (2008) 094502}
  [\href{https://arxiv.org/abs/0705.4295}{{\ttfamily 0705.4295}}].

\bibitem{radpseudo}
A.V.~Radyushkin, \emph{{Quasi-parton distribution functions, momentum
  distributions, and pseudo-parton distribution functions}},
  \href{https://doi.org/10.1103/PhysRevD.96.034025}{\emph{Phys. Rev. D}
  {\bfseries 96} (2017) 034025}
  [\href{https://arxiv.org/abs/1705.01488}{{\ttfamily 1705.01488}}].

\bibitem{jiquasi}
X.~Ji, \emph{{Parton Physics on a Euclidean Lattice}},
  \href{https://doi.org/10.1103/PhysRevLett.110.262002}{\emph{Phys. Rev. Lett.}
  {\bfseries 110} (2013) 262002}
  [\href{https://arxiv.org/abs/1305.1539}{{\ttfamily 1305.1539}}].

\bibitem{pionquasi}
J.-W.~Chen, H.-W.~Lin and J.-H.~Zhang, \emph{{Pion generalized parton
  distribution from lattice {QCD}}},
  \href{https://doi.org/10.1016/j.nuclphysb.2020.114940}{\emph{Nucl. Phys. B}
  {\bfseries 952} (2020) 114940}
  [\href{https://arxiv.org/abs/1904.12376}{{\ttfamily 1904.12376}}].

\bibitem{nucleonquasi1}
H.-W.~Lin, \emph{{Nucleon Tomography and Generalized Parton Distribution at
  Physical Pion Mass from Lattice QCD}},
  \href{https://doi.org/10.1103/PhysRevLett.127.182001}{\emph{Phys. Rev. Lett.}
  {\bfseries 127} (2021) 182001}
  [\href{https://arxiv.org/abs/2008.12474}{{\ttfamily 2008.12474}}].

\bibitem{nucleonquasi2}
C.~Alexandrou, K.~Cichy, M.~Constantinou, K.~Hadjiyiannakou, K.~Jansen,
  A.~Scapellato et~al., \emph{{Unpolarized and helicity generalized parton
  distributions of the proton within lattice {QCD}}},
  \href{https://doi.org/10.1103/PhysRevLett.125.262001}{\emph{Phys. Rev. Lett.}
  {\bfseries 125} (2020) 262001}
  [\href{https://arxiv.org/abs/2008.10573}{{\ttfamily 2008.10573}}].

\bibitem{fwdletter}
A.J.~Chambers, R.~Horsley, Y.~Nakamura, H.~Perlt, P.E.L.~Rakow, G.~Schierholz
  et~al., \emph{{Nucleon Structure Functions from Operator Product Expansion on
  the Lattice}},
  \href{https://doi.org/10.1103/PhysRevLett.118.242001}{\emph{Phys. Rev. Lett.}
  {\bfseries 118} (2017) 242001}
  [\href{https://arxiv.org/abs/1703.01153}{{\ttfamily 1703.01153}}].

\bibitem{comptonproceedings}
A.~Hannaford-Gunn, R.~Horsley, Y.~Nakamura, H.~Perlt, P.E.L.~Rakow,
  G.~Schierholz et~al., \emph{{Scaling and higher twist in the nucleon Compton
  amplitude}}, \href{https://doi.org/10.22323/1.363.0278}{\emph{PoS} {\bfseries
  LATTICE2019} (2020) 278} [\href{https://arxiv.org/abs/2001.05090}{{\ttfamily
  2001.05090}}].

\bibitem{fwdpaper}
K.U.~Can, A.~Hannaford-Gunn, R.~Horsley, Y.~Nakamura, H.~Perlt, P.E.L.~Rakow
  et~al., \emph{{Lattice QCD evaluation of the Compton amplitude employing the
  Feynman-Hellmann theorem}},
  \href{https://doi.org/10.1103/PhysRevD.102.114505}{\emph{Phys. Rev. D}
  {\bfseries 102} (2020) 114505}
  [\href{https://arxiv.org/abs/2007.01523}{{\ttfamily 2007.01523}}].

\bibitem{utkucallangross}
K.U.~Can, A.~Hannaford-Gunn, E.~Sankey, R.~Horsley, Y.~Nakamura, H.~Perlt
  et~al., \emph{{Investigating the low moments of the nucleon structure
  functions in lattice QCD}},  in \emph{{38th International Symposium on
  Lattice Field Theory}}, 10, 2021
  [\href{https://arxiv.org/abs/2110.01310}{{\ttfamily 2110.01310}}].

\bibitem{interlacingsubtraction}
A.~Hannaford-Gunn and E.~Sankey, \emph{{Investigating the Compton amplitude
  subtraction function in lattice QCD}},  in \emph{{38th International
  Symposium on Lattice Field Theory}}, 12, 2021.

\bibitem{collabffs}
A.~Chambers, J.~Dragos, R.~Horsley, Y.~Nakamura, H.~Perlt, D.~Pleiter et~al.,
  \emph{{Electromagnetic form factors at large momenta from lattice QCD}},
  \href{https://doi.org/10.1103/PhysRevD.96.114509}{\emph{Phys. Rev. D}
  {\bfseries 96} (2017) 114509}
  [\href{https://arxiv.org/abs/1702.01513}{{\ttfamily 1702.01513}}].

\bibitem{configs}
W.~Bietenholz, V.~Bornyakov, M.~Göckeler, R.~Horsley, W.G.~Lockhart,
  Y.~Nakamura et~al., \emph{{Flavour blindness and patterns of flavour symmetry
  breaking in lattice simulations of up, down and strange quarks}},
  \href{https://doi.org/10.1103/PhysRevD.84.054509}{\emph{Phys. Rev. D}
  {\bfseries 84} (2011) 054509}
  [\href{https://arxiv.org/abs/1102.5300}{{\ttfamily 1102.5300}}].

\bibitem{offfwdpaper}
A.~Hannaford-Gunn, K.U.~Can, R.~Horsley, Y.~Nakamura, H.~Perlt, P.E.L.~Rakow
  et~al., \emph{{Generalized parton distributions from the off-forward Compton
  amplitude in lattice QCD}},
  \href{https://doi.org/10.1103/PhysRevD.105.014502}{\emph{Phys. Rev. D}
  {\bfseries 105} (2022) 014502}
  [\href{https://arxiv.org/abs/2110.11532}{{\ttfamily 2110.11532}}].

\bibitem{pymc3_no1}
J.~Salvatier, T.V.~Wiecki and C.~Fonnesbeck, \emph{{Probabilistic programming
  in Python using PyMC3}}, {\emph{PeerJ Computer Science} {\bfseries 2:e55}
  (2016) } [\href{https://arxiv.org/abs/1507.08050}{{\ttfamily 1507.08050}}].

\bibitem{pymc3_no2}
M.D.~Hoffman and A.~Gelman, \emph{{The No-U-Turn Sampler: Adaptively Setting
  Path Lengths in Hamiltonian Monte Carlo}}, {\emph{Journal of Machine Learning
  Research} {\bfseries 15} (2014) 1593}
  [\href{https://arxiv.org/abs/1111.4246}{{\ttfamily 1111.4246}}].

\bibitem{Pobylitsa_2002}
P.V.~Pobylitsa, \emph{{Disentangling positivity constraints for generalized
  parton distributions}},
  \href{https://doi.org/10.1103/PhysRevD.65.114015}{\emph{Phys. Rev. D}
  {\bfseries 65} (2002) 114015}
  [\href{https://arxiv.org/abs/hep-ph/0201030}{{\ttfamily hep-ph/0201030}}].

\bibitem{Pobylitsa_2002_IPS}
P.V.~Pobylitsa, \emph{{Positivity bounds on generalized parton distributions in
  impact parameter representation}},
  \href{https://doi.org/10.1103/PhysRevD.66.094002}{\emph{Phys. Rev. D}
  {\bfseries 66} (2002) 094002}
  [\href{https://arxiv.org/abs/hep-ph/0204337}{{\ttfamily hep-ph/0204337}}].

\bibitem{Pobylitsa_2004}
P.V.~Pobylitsa, \emph{{Virtual Compton scattering in the generalized Bjorken
  region and positivity bounds on generalized parton distributions}},
  \href{https://doi.org/10.1103/PhysRevD.70.034004}{\emph{Phys. Rev. D}
  {\bfseries 70} (2004) 034004}
  [\href{https://arxiv.org/abs/hep-ph/0211160}{{\ttfamily hep-ph/0211160}}].

\bibitem{compton_positivity}
A.~De~R\'ujula, \emph{An introduction to the positivity constraints on
  absorptive amplitudes and their experimental implications},  in \emph{{7th
  Rencontres de Moriond}: {multiparticle phenomena and inclusive reactions}},
  p.~405, 1972,
  \href{https://inspirehep.net/literature/1519926}{https://inspirehep.net/literature/1519926}.

\bibitem{montecarlofit}
R.~Horsley, Y.~Nakamura, H.~Perlt, P.E.L.~Rakow, G.~Schierholz, K.~Somfleth
  et~al., \emph{{Structure functions from the Compton amplitude}},
  \href{https://doi.org/10.22323/1.363.0137}{\emph{PoS} {\bfseries LATTICE2019}
  (2020) 137} [\href{https://arxiv.org/abs/2001.05366}{{\ttfamily
  2001.05366}}].

\bibitem{Haar:2017ubh}
T.R.~Haar, Y.~Nakamura and H.~St{\"u}ben, \emph{{An update on the B{QCD} Hybrid
  Monte Carlo program}},
  \href{https://doi.org/10.1051/epjconf/201817514011}{\emph{EPJ Web Conf.}
  {\bfseries 175} (2018) 14011}
  [\href{https://arxiv.org/abs/1711.03836}{{\ttfamily 1711.03836}}].

\bibitem{Edwards:2004sx}
{\scshape SciDAC Collaboration, LHPC Collaboration, UK{QCD} Collaboration}
  collaboration, \emph{{The Chroma software system for lattice {QCD}}},
  \href{https://doi.org/10.1016/j.nuclphysbps.2004.11.254}{\emph{Nucl.Phys.Proc.Suppl.}
  {\bfseries 140} (2005) 832}
  [\href{https://arxiv.org/abs/hep-lat/0409003}{{\ttfamily hep-lat/0409003}}].

\end{thebibliography}\endgroup
